# Macho Proper Motions From Optical/Infrared Photometry

**Andrew Gould**[1]

Dept of Astronomy, Ohio State University, Columbus, OH 43210

**Douglas L. Welch**

Dept of Physics & Astronomy, McMaster University

Hamilton, ON L8S 4M1 Canada

gould@payne.mps.ohio-state.edu; welch@physics.mcmaster.ca

## Abstract

Optical/infrared photometry can double the number of proper motion measurements of Massive Compact Objects (MACHOs) relative to single band photometry. The proper motion of a MACHO can be measured by finding the ratio $q$ of the (known) radius of the source star to the Einstein radius of the MACHO, $q = \theta_s/\theta_e$. A classic method for doing this is to look for the effect on the light curve of the finite size of the source. A modification of this method proposed by Witt (1995) is to look for color changes in the light curve due to the fact that the limb darkening of the source is different in different bands. We demonstrate that the "classical" method is not feasible unless the MACHO actually transits the source: if the MACHO passes at say 1.5 source radii, there is still a sizable $\sim 5\%$ effect, but the light curve cannot be distinguished from point-source light curves with different parameters. However, color measurements in $V$ ($0.55\,\mu$m) and $H$ ($1.65\,\mu$m) reduce the errors by a factor $\sim 120$ and permit proper motion measurements at impact parameters of up to 2 source radii. Color maps in $V-H$ are also useful in the detection of planetary systems. Giant stars have a "red ring" in such maps. A planet which transits this ring gives rise to a distinctive signature which can help in the measurement of the planetary system's proper motion.

1 Alfred P. Sloan Foundation Fellow

# 1. Introduction

The MACHO collaboration (Alcock et al. 1995; Bennett et al. 1995) and the OGLE collaboration (Udalski et al. 1994) have found a total of 56 candidate Massive Compact Object (MACHO) events toward sources in the Galactic bulge. The events are detected by measuring the flux of the source star as a function of time. The flux is magnified by a factor $A[x(t)]$ (Paczyński 1986)

$$A(x) = \frac{x^2+2}{x(x^2+4)^{1/2}}, \tag{1.1}$$

where $x$ is the separation between the lens and source in units of the Einstein radius, $\theta_e$,

$$\theta_e \equiv \sqrt{\frac{4GMD_{\rm LS}}{c^2 D_{\rm OL} D_{\rm OS}}}, \tag{1.2}$$

$M$ is the mass of the lens, and $D_{\rm OL}$, $D_{\rm LS}$, and $D_{\rm OS}$ are the distances between the observer, lens, and source. If the lens is moving with speed $v$ relative to the observer-source line of sight, then

$$x(t) = \sqrt{\omega^2(t-t_0)^2+\beta^2}, \tag{1.3}$$

where $t_0$ is the time of maximum magnification, $\beta$ is the impact parameter in units of $\theta_e$, and $\omega^{-1}$ is the Einstein ring crossing time,

$$\omega^{-1} = \frac{\theta_e D_{\rm OL}}{v}. \tag{1.4}$$

The only information that one recovers from the event about the lens itself is the time scale, which is a complicated combination of the three physical parameters that one would like to know, $M$, $D_{\rm OL}$, and $v$. (The distance to the source is assumed known from spectroscopic parallax.) In order to understand the nature of these events, one would like to gain additional information. If one could measure the "proper motion", $\mu$,

$$\mu = \omega\theta_e, \tag{1.5}$$

and the "parallax" (from which one determines the projected speed $\tilde{v} = [D_{\rm OS}/D_{\rm LS}]v$), one could obtain individual masses, distances, and speeds of the objects (Gould 1992, 1995c). One parallax (of an unusually long event, $\omega^{-1} \sim 4\,\text{months}$) has been measured from the ground (Bennett et al. 1995), and parallaxes of lensing events with bulge giant sources could be measured routinely using a satellite telescope (Gould 1994b, 1995a).

A number of methods have been advanced for measuring proper motions. One method is based on the fact that if the lens transits (or nearly transits) the face of the source, then the light curve deviates from standard form (1.1) (Gould 1994a; Nemiroff & Wickramasinghe 1994; Witt & Mao 1994). Other methods have been advanced by Maoz & Gould (1994), Hog, Novikov, & Polnarev (1994), Simmons, Willis, & Newsam (1995), Loeb & Sasselov (1995), and Han, Narayanan, & Gould (1995).

The most widely applicable of all of these methods is the photometry of transit events. For giant sources, the mean source radius is $\langle r_s\rangle \sim 22\,r_\odot$ (Gould 1995b). For events generated by bulge lenses, the Einstein ring (projected on the source plane) is $\sim 400\,r_\odot(M/0.4\,M_\odot)^{1/2}(D_{\rm LS}/\text{kpc})^{1/2}$. Hence of order 5% of these events are transits. In addition to being big, bulge giants are ideal sources because they are bright and hence can be easily and accurately photometered. In addition, as mentioned above, a future satellite could measure parallaxes for giants.

Even if the lens does not transit the star, the light curve is still affected. As we will show below, the magnification is then

$$\tilde{A}_q(x) = A(x)\biggl(1 + \frac{1}{8}\Lambda\frac{q^2}{x^2}\biggr), \tag{1.6}$$

where

$$q = \frac{\theta_s}{\theta_e}, \tag{1.7}$$

$\theta_s$ is the angular radius of the source, and $\Lambda \sim 1$ is a factor that depends on the limb darkening of the source. At $x = 1.5q$ for example, the flux is augmented by $\sim 5\%$. Gould (1994a) argued, however, that this effect would be essentially undetectable because it could masquerade as a change in the other parameters. Only if the lens actually transited the source would there be enough structure in the light curve to detect the finite size of the source.

Witt (1995) pointed out that limb darkening would actually generate color terms in the light curve. The sense of the effect of limb-darkening is that for redder wavelengths, the star will more nearly resemble a uniform disk. This implies that a $V - H$ color map of a star has a bright red ring. Hence the flux increment in equation (1.6) is greater when the star is observed in redder bands. Since gravitational lensing is usually achromatic, this color term is an unambiguous signature that the lens is resolving the source.

Here we confirm the claim by Gould (1994a) that the deviation of the light curve cannot be detected in a single band when $\beta > q$. But we also show that two-band (specifically $V$ and $H$) photometry increases the sensitivity by a factor $\sim 120$, corresponding to an increased statistical power $\sim 10^4$. Optical/infrared photometry enables the detection of proper motions even for $\beta \sim 2q$, thus about doubling the number of proper motion measurements relative to single-band photometry.

## 2. Magnification of Limb-Darkened Stars

Consider a ring of light of radius $y\theta_e$ whose center is separated by $x\theta_e$ from the lens, with $x \ll 1$. A point at $\phi$ along the ring will be separated from the lens by $r = (x^2 + y^2 - 2xy\cos\phi)^{1/2}$. Since $r \ll 1$, $A(r) \to r^{-1}$. Hence the ring will be magnified by

$$\tilde{A}(x,y) = \frac{1}{2\pi}\int d\phi (x^2 + y^2 - 2xy\cos\phi)^{-1/2} = A(x)\left[1 + \frac{1}{4}\left(\frac{y}{x}\right)^2 + \frac{9}{64}\left(\frac{y}{x}\right)^4 + \ldots\right]. \tag{2.1}$$

Limb darkening of the surface brightness $S$ of stars can be parameterized by

$$\frac{S(\theta)}{S(0)} = 1 - \kappa_1 Y - \kappa_2 Y^2, \qquad Y \equiv 1 - \sqrt{1 - \frac{\theta^2}{\theta_s^2}}. \tag{2.2}$$

Limb-darkening coefficients for a cool (4500 K) giant ($\log g = 1.5$) of solar metallicity in $V$ and $H$ are $\kappa_1^V = 0.798$, $\kappa_2^V = -0.007$, $\kappa_1^H = 0.206$, and $\kappa_2^H = 0.331$ (Manduca, Bell & Gustafsson 1977; Manduca 1979). Integrating the flux over the entire star [and keeping terms in equation (2.1) only to second order] yields a total magnification given by equation (1.6), with

$$\Lambda = \frac{30 - 14\kappa_1 - 8\kappa_2}{30 - 10\kappa_1 - 5\kappa_2}. \tag{2.3}$$

Note that for $q \ll 1$ equation (2.3) remains valid even for $x \gtrsim 1$ (Gould 1994a). For $V$ and $H$ the limb-darkening corrections are $\Lambda^V = 0.86$ and $\Lambda^H = 0.93$. (For $U$, $B$, $V$, $I$, $H$, and $K$, $\Lambda$ is respectively 0.77, 0.82, 0.86, 0.89, 0.91, 0.93, and 0.95. In this paper we adopt $V$ and $H$ as the "blue" and "red" bands in order to obtain the largest leverage in $\Lambda^{\rm blue} - \Lambda^{\rm red}$ while meeting practical observational constraints. We choose $H$ rather than $K$ to avoid the greater sky noise in $K$. We choose $V$ rather than $U$ or $B$, first because during bright time most bulge giants are fainter than the sky in the latter bands, and second because many bulge fields are heavily obscured. We note nevertheless that for some fields and some observing conditions, other combinations of colors may be better.)

## 3. Statistics of Light-Curve Fitting

The flux from a lensed star of finite size can be written as a function of six parameters $a_i = (t_0, \beta, \omega, F_0, B, \epsilon)$

$$F(t) = F_0 A[x(t; t_0, \beta, \omega)]\left(1 + \frac{\epsilon}{x^2}\right) + B, \tag{3.1}$$

where $F_0$ is the flux from the source without lensing, $B$ is the light from any unresolved and unlensed star such as the lens itself or a binary companion to the source, and $\epsilon \equiv \Lambda q^2/8$. Note that $B$ must be included as a free parameter even if the best fit is consistent with no unlensed sources because such sources cannot be ruled out *a priori*. For simplicity of exposition we set the zero point of time to $t_0 = 0$ and normalize the unit of time so that $\omega = 1$ and the unit of flux so that $F_0 = 1$. We assume that the best fit value of $B \ll 1$, and that $\epsilon \ll 1$. Suppose that a series measurements of $F(t_k)$ are made at times $t_k$, with uncertainties $\sigma_k$, and that equation (3.1) is fit to the measurements. The covariance matrix of the errors of this fit is given by $c_{ij}$ where

$$c = b^{-1} \qquad b_{ij} = \sum_k \sigma_k^{-2} \frac{\partial F(t_k)}{\partial a_i} \frac{\partial F(t_k)}{\partial a_j}. \tag{3.2}$$

We evaluate $c_{ij}$ for $\beta = 0.1$ assuming that measurements are made at a rate $N\omega$ with (photon-limited) precision $\sigma = \sigma_0 A^{1/2}$ from $t = t_0 - \omega^{-1}$ (when the event is first noticed) until $t = t_0 + 5\omega^{-1}$. We find

$$\sqrt{c_{ii}} = N^{-1/2}\frac{\sigma_0}{0.01}\left(\, 0.0013 \quad 3.0056 \quad 0.2683 \quad 0.0867 \quad 0.0963 \quad 0.2978 \,\right),$$

$$\tilde{c}_{ij} = \begin{pmatrix} 1 & -0.015109 & -0.012763 & -0.000661 & 0.002785 & -0.015135 \\ -0.015109 & 1 & 0.985820 & 0.671057 & -0.772565 & 0.999998 \\ -0.012763 & 0.985820 & 1 & 0.784270 & -0.865561 & 0.985477 \\ -0.000661 & 0.671057 & 0.784270 & 1 & -0.987385 & 0.669518 \\ 0.002785 & -0.772565 & -0.865561 & -0.987385 & 1 & -0.771250 \\ -0.015135 & 0.999998 & 0.985477 & 0.669518 & -0.771250 & 1 \end{pmatrix}, \tag{3.3}$$

where $\tilde{c}_{ij} = c_{ij}/\sqrt{c_{ii}c_{jj}}$ is the matrix of correlation coefficients, and where the indices refer to the vector components $(t_0, \beta, \omega, F_0, B, \epsilon)$. The measurement error of $\epsilon$ is very large, corresponding to a fractional error in $q$

$$\frac{\delta q}{q} = 4\frac{\sqrt{c_{\epsilon,\epsilon}}}{\Lambda q^2} \sim \frac{120}{\sqrt{N}}\frac{\sigma_0}{0.01}\left(\frac{\beta}{q}\right)^2, \tag{3.4}$$

Clearly when $\beta > q$ (the regime for which this calculation is valid), $q$ cannot be measured unless the photometry is extraordinary.

The reason for the poor precision is the extreme degeneracy of $\epsilon$ with the other parameters, most notably with $\beta$ for which $1 - \tilde{c}_{\beta,\epsilon} = 2 \times 10^{-6}$. That is, the finite-source light curve ($\epsilon > 0$) is very well mimicked by a point-source light curve with parameters adjusted by

$$\delta a_i = -\epsilon\frac{c_{i,\epsilon}}{c_{\epsilon,\epsilon}}. \tag{3.5}$$

If the $a_i$ are held fixed for $i \neq 6$, then the light curves with and without the finite-source term are easily distinguished. But, if the remaining parameters are adjusted by equation (3.5), the curves are essentially the same. See Figure 1.

Let us now imagine that the observations are carried out simultaneously in $V$ and $H$ with, for example, an optical/infrared camera equipped with a dichroic beam splitter. For simplicity we assume that the errors are equal in the two bands. If the $V$ and $H$ light curves are initially fit separately, there will be 12 parameters, two for each of the parameters in the single band case. The covariance matrix will then be block diagonal, with each $6 \times 6$ block a duplicate of equation (3.3). However, we may now impose the following four constraints:

$$t_{0,V} = t_{0,H}, \quad \beta_V = \beta_H, \quad \omega_V = \omega_H, \quad 0.93\epsilon_V = 0.86\epsilon_H. \tag{3.6}$$

The last constraint arises because the finite-size term differs in the two bands by a known amount. As a practical matter, each constraint is imposed by successively

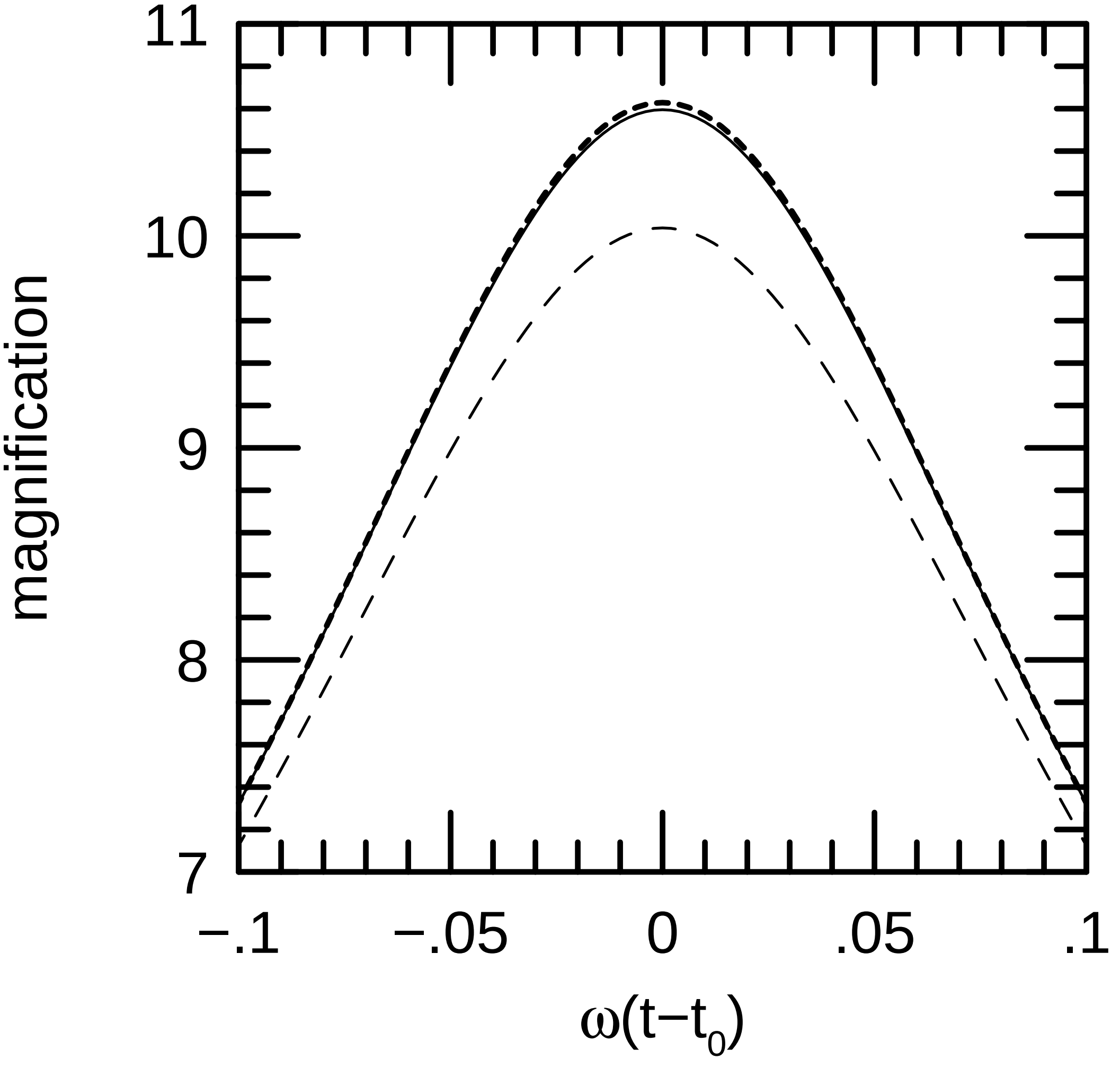

Figure 1. Theoretical light curves for a lensing event with finite source size $q = 0.067$ and $\beta = 0.1$. Shown are the true curve (solid), the best fit curve assuming a point source but with the remaining five parameters freely adjusted (short dashes), and the point-source curve with the remaining parameters held fixed (long dashes). Note that although the true curve can be easily distinguished from the point-source curve with fixed parameters, it is indistinguishable from the best-fit point-source curve.

adjusting $c_{ij}$ by

$$c_{ij} \rightarrow c_{ij} + \frac{c_{ik}\alpha_k c_{jl}\alpha_l}{\alpha_m c_{mn}\alpha_n}, \tag{3.7}$$

where $\alpha_i a_i = 0$ is the constraint. For example, the last constraint is represented by $\alpha_i = (0, 0, 0, 0, 0, 0.93, 0, 0, 0, 0, 0, -0.86)$. Carrying out this procedure, we then find that the covariance matrix is given by

$$\sqrt{c_{ii}} = N^{-1/2}\frac{\sigma_0}{0.01}\,(\,0.0009 \quad 0.0264 \quad 0.0323 \quad 0.0462 \quad 0.0445 \quad 0.0025\,),$$

$$\tilde{c}_{ij} = \begin{pmatrix} 1 & 0.002 & 0.013 & 0.013 & -0.014 & 0.000 \\ 0.002 & 1 & 0.232 & 0.320 & -0.351 & 0.986 \\ 0.013 & 0.232 & 1 & 0.981 & -0.962 & 0.070 \\ 0.013 & 0.320 & 0.981 & 1 & -0.993 & 0.159 \\ -0.014 & -0.351 & -0.962 & -0.993 & 1 & -0.193 \\ 0.000 & 0.986 & 0.070 & 0.159 & -0.193 & 1 \end{pmatrix}, \qquad (3.8)$$

Note that the error in $\epsilon$ has fallen by a factor $\sim 120$ relative to equation (3.3). That is, the "120" in equation (3.4) is replaced by unity. Thus for a lens that has an impact parameter of two stellar radii ($\beta = 2q$), $q$ can be detected at the $3\,\sigma$ level with $N = 150$ and $\sigma_0 = 1\%$. The mathematical reason for this enormous improvement is that while the light curve in each band can be well fit with a point-source curve by suitably adjusting the remaining parameters according to equation (3.5), these adjustments are different for $V$ and $H$. Hence, the adjustments of the other parameters cannot simulate the change in both bands simultaneously. From a physical standpoint, the finite source term introduces color terms which cannot be mimicked by ordinary lensing of a point source, because lensing is achromatic.

This last point deserves closer examination. In fact, ordinary lensing can have color terms if the source is a blend, $B \neq 0$. If for example the lensed source is bluer than the unlensed blended light, then the event will become bluer as it approaches the peak. However, strong constraints can be obtained on the colors of the lensed and unlensed sources by comparing the color of the event at $x \gg 1$ and at $x \sim 0.5$. Note that this comparison is not significantly affected by the finite size of the source because typically $q \lesssim 0.1$. Once the colors of the the lensed and unlensed sources are known, it is trivial to predict the expected color at the peak for a point-source event. Deviations from this prediction then must be due to a finite source.

Of course, the fitting procedure automatically takes all this into account, but the physical explanation is nonetheless useful. We showed that $N = 150$ measurements per Einstein crossing time are necessary to detect $q$ at $q = 2\beta$. This implies a total of 900 measurements during the time interval $-\omega^{-1} < t < 5\omega^{-1}$. But according to the above analysis, not all of these measurements are equally necessary. To test this, we imagine that the measurements are made at $N = 150\omega$ for 0.5 crossing times around the peak, around $x = 0.5$, and around $x = 4.7$, and that during the remaining time the observations are conducted at only $N = 15\omega$. We find that this reduction from 900 to 290 measurements increases the errors by only $\sim 40\%$.

In our illustrative example, we chose $\beta = 0.1$. How does sensitivity to $q$ depend on $\beta$? To address this question, we write

$$\frac{\delta q}{q} = \frac{\eta}{\sqrt{N}}\frac{\sigma_0}{0.01}\left(\frac{\beta}{q}\right)^2, \tag{3.9}$$

and evaluate $\eta$ for various values of $\beta$. We find for $\beta = 0.05,\ 0.10,\ 0.15,\ 0.20$, and 0.3, that $\eta = 0.54,\ 0.98,\ 1.26,\ 1.43$, and 1.84. That is, color measurements of proper motions depend mainly on the number of stellar radii of the impact parameter $\beta/q$, and only weakly on $\beta$ and $q$ separately.

We conclude that measuring proper motions of lensing events from optical/infrared colors is substantially more effective than similar measurements in a single band, which latter are possible only for transits. The rate at which proper motions can be conclusively detected using a combination of optical and infrared photometry is at least twice the single bandpass rate.

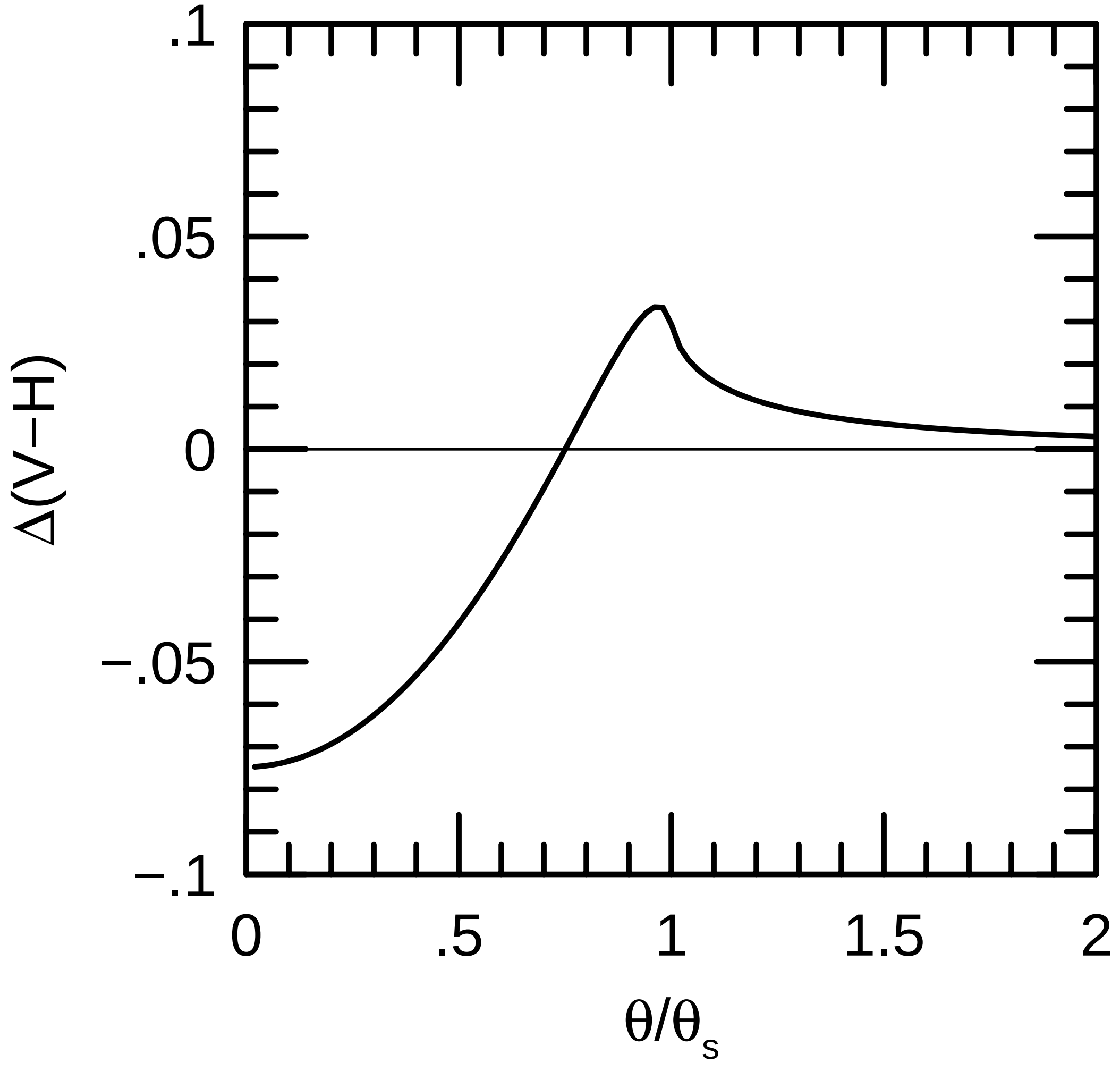

Figure 2. Change in color as a function of source-lens separation in units of the stellar radius, assuming $\theta_e \gg \theta_s$. The zero point of the color is that of the unlensed source.

## 4. Transits and Planetary System Lenses

If the lens transits the face of the star, then the light curve in each band deviates substantially from the limiting form of equation (1.6). In this case it is straight forward to measure the proper motion from each band separately. Colors are not needed. However, the color term does help confirm the transit. We show this color (relative to the color of the unlensed star) as a function of source-lens separation in Figure 2.

The main value of colors during transits is in the search for planetary systems. If a star with a small planet having mass $m \ll M$ acts as a lens, then for most of the lensing event, the light curve will look like a standard microlensing light curve. However, as first pointed out by Mao & Paczyński (1991) in some cases the planet can dramatically alter the light curve during a fraction $(m/M)^{1/2}$ of the Einstein crossing time. Gould & Loeb (1992) analyzed the problem in detail but restricted their attention to the case of point sources. They estimated for planets in the "lensing zone" [with projected separations from the parent star $a \sim$ 3–6 AU $(M/M_\odot)^{1/2}$] that the planet would give rise to a perturbation $\gtrsim 5\%$ in $\sim 17\%(m/m_{\rm jupiter})^{1/2}$ of the events. Note that the Einstein radius associated with a Jupiter mass planet is $\sim 50\, r_\odot/R_0$ where $r_\odot$ is the radius of the Sun and $R_0 \sim 8\,$kpc is the galactocentric distance. Hence the point-source approximation remains valid even for bulge giant sources with radius $r \sim 10\, r_\odot$.

However, for smaller planets, particularly Earth mass planets, the point-source approximation breaks down. In this case, one can detect the planet only if the caustic associated with the planet actually transits the star. Even then the degree of brightening may be quite modest. One can show that if the planet is separated from the star by more than one Einstein radius, then the star brightens by a factor $A \sim 1 + 2(m/M)(\theta_e/\theta_s)^2$. For an Earth-mass planet at 2 kpc from the Galactic center, this is $A \sim 1 + 0.02(r/20r_\odot)^2$. The effect is therefore detectable even for large sources. This is important because the probability that the planet will transit the star $\sim (r/R_0)/\theta_e$. The mean radius of giants ($M_I < 0.5$) in the Galactic bulge is $\langle r\rangle \sim 22\, r_\odot$ (Gould 1995b). The probability that a planet will transit such a star is $\sim 2\%(M/M_\odot)^{1/2}$. Given that an aggressive search could detect $\gtrsim 100$ giant-source events per year, there is a significant chance to detect some small planets.

If such planets are detected, one would like to know as much about them as possible. The peak excess light $A-1$ constrains the quantity $(m/M)(\theta_e/\theta_s)^2$. If one knew the geometry of the transit (i.e., the impact parameter in units of the source radius) then the duration of the transit would yield $(\theta_s/\theta_e)$ and hence the proper motion $\mu = \omega\theta_e = v/D_{\rm OL}$. The question is, how does one know the geometry of the transit? In principal, this can be extracted from single band photometry by measuring the length of time it takes the caustic to move across the edge of the star, that is the time from no excess magnification to full excess. This [together with the above mentioned measurement of $(m/M)^{1/2}\theta_e$] gives the angle between the edge of the star and the trajectory of the lens. In practice, much more information would be available if the small lens were transiting a narrow ring rather than the edge of a disk. The enhanced sensitivity of a ring has been pointed out by Loeb & Sasselov (1995) who showed that such a ring exists in narrow band CaII images. As shown in Figure 3, $V - H$ color maps also have a narrow ring. As with CaII images, this ring can help resolve the proper motion of planetary systems. Under most observing conditions $V - H$ color maps will have substantially higher signal to noise than CaII images in part because they are based on broad-band photometry and in part because $V$ band is less heavily extincted than the 393 nm CaII line.

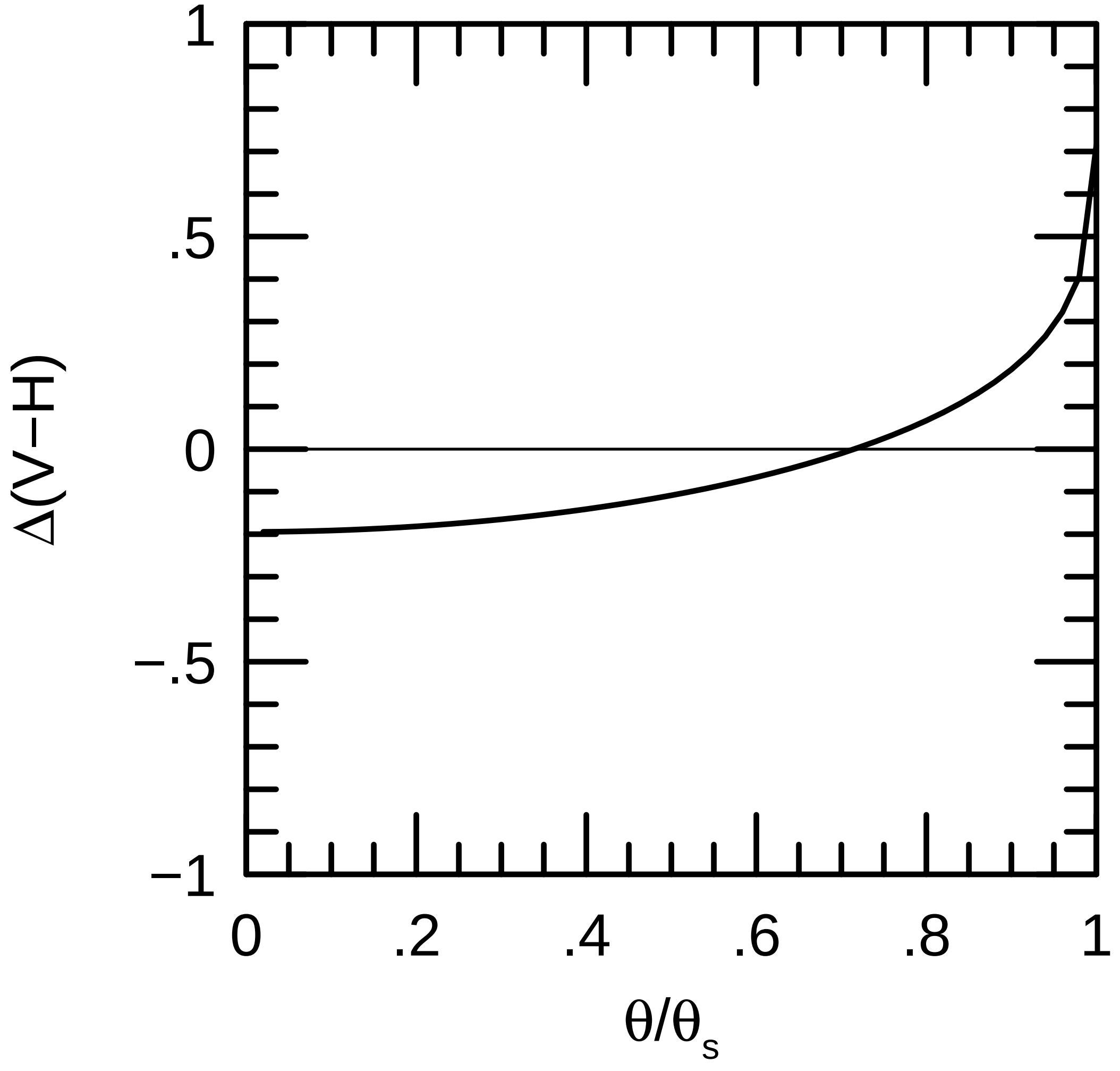

Figure 3. Change in $V-H$ color of the *excess light* $(A-1)$ due to a very small lens passing over the face of the star, $\theta_e \ll \theta_s$. The zero point of color is that of the unlensed star. Note that there is a large color shift for $0.9 \lesssim \theta/\theta_s \leq 1$ which gives rise to a "red ring" in the color map. This ring is helpful in the analysis of events generated by planetary systems with small (e.g. Earth-mass) planets.

**Acknowledgements**: We would like to thank D. DePoy for stimulating discussions. Work was supported in part by a grant to AG from the NSF AST 94-20746 and by a research grant to DLW from the Natural Science and Engineering Research Council (NSERC) of Canada.